\documentclass[aps,prx,twocolumn,notitlepage,superscriptaddress]{revtex4-2}
\usepackage{bm} \usepackage{graphicx} \usepackage{amsmath,amsthm,stmaryrd,amssymb} 
\usepackage{color}
\usepackage{hyperref}

\usepackage{algorithm}
\usepackage{algpseudocode}

\DeclareMathOperator*{\Tr}{Tr}

\newcommand{\ket}[1]{\vert{#1}\rangle}
\newcommand{\bra}[1]{\langle{#1}\vert}

\makeatletter
\newcommand{\StatexIndent}[1][3]{%
  \setlength\@tempdima{\algorithmicindent}%
  \Statex\hskip\dimexpr#1\@tempdima\relax}
\makeatother

\begin{document}

\title{Accuracy vs Memory Advantage in the Quantum Simulation of Stochastic 
Processes}

\author{Leonardo Banchi} \email{leonardo.banchi@unifi.it}
\affiliation{Department of Physics and Astronomy, University of Florence,
via G. Sansone 1, I-50019 Sesto Fiorentino (FI), Italy}
\affiliation{ INFN Sezione di Firenze, via G. Sansone 1, I-50019, Sesto Fiorentino (FI), Italy }

\begin{abstract}
	Many inference scenarios rely on extracting relevant information from known
	data in order to make future predictions. When the underlying stochastic
	process satisfies certain assumptions, there is a direct mapping between
	its exact classical and quantum simulators, with the latter asymptotically
	using less memory. 
	Here we focus on studying whether such quantum advantage persists when
	those assumptions are not satisfied, and the model is doomed to have
	imperfect accuracy. By studying the trade-off between accuracy and 
	memory requirements, we show that quantum models can reach the same 
	accuracy with less memory, or alternatively, better accuracy with the 
	same memory. Finally, we discuss the implications of this result for 
	learning tasks. 
\end{abstract}

\maketitle

\section{Introduction}

The ability to learn from experience and to make predictions about possible future outcomes is crucial 
in all quantitative sciences. Since humans and machines have a limited amount of memory, 
learning complex processes requires distilling and storing only the relevant information from the training data 
that is useful for making predictions. 
For temporal data, the current state-of-the-art in classical machine learning is based on the transformers 
architecture  \cite{vaswani2017attention}, which uses the self-attention mechanism to dynamically focus on  
what is relevant in the data stream. Such an architecture culminates a series of tweaks performed by the 
machine learning community over the last decades to solve practical problems, making it difficult to extract the 
underlying mathematical principles, in spite of some progresses \cite{phuong2022formal}. 
On the other hand, stochastic processes called $\varepsilon$-machines have been developed
on rigorous mathematical grounds, to formally define what 
past information a learner needs to store for predicting future outcomes 
\cite{crutchfield1989inferring,crutchfield1997statistical,shalizi2001computational}. 
However, numerically fitting the model from data is more complicated \cite{shalizi2002algorithm}. 

From a quantum information perspective, 
whenever a stochastic process can be {\it exactly} expressed as an
$\varepsilon$-machine with a finite amount of memory, it has been shown that a
unitary quantum simulator 
is capable of exactly simulating the same process, asymptotically storing less memory 
\cite{gu2012quantum,elliott2020extreme,elliott2021memory,elliott2021quantum}. 
In other terms, quantum advantage in memory use can be achieved by 
using quantum states and probability amplitudes, 
which provide other sources of stochasticity when the 
memory states are not orthogonal. Moreover, 
this advantage has also been observed in tensor network simulations 
\cite{glasser2019expressive,yang2018matrix,yang2023provably}. 

In this work we study whether the quantum advantage persists when we relax the requirement of exact 
simulation. Real-world datasets, in general, cannot be exactly modelled as an $\varepsilon$-machine with 
a given amount of memory. Therefore, it is unclear whether we can formally expect quantum advantage with them. 
Nonetheless,
starting from  a generic stochastic process, it is reasonable to 
assume that there exists a larger $\varepsilon$-machine, possibly with infinite memory, 
capable of exactly modelling the data.  The latter can be exactly expressed
as a unitary quantum simulator, which typically 
shows memory advantage and never requires more memory \cite{elliott2021memory}. 
Does the advantage persist in constrained memory scenarios? 
If this were the case, 
quantum simulators of real-world stochastic processes could 
reach the same accuracy of classical ones with less memory, 
or, alternatively, achieve better accuracies with the same amount of memory.
Moreover, since the information contained in the quantum memory constraints the generalization 
error \cite{banchi2021generalization,banchi2023statistical}, it is tempting to expect that a
memory advantage results in the ability to learn the model with less data.

Motivated by the above questions, and given the difficulty in extracting general predictions from 
real-world data, we focus on toy problems
that are easier to interpret and model.  Instead of the mentioned bottom-up construction,
where, given a real-world problem and a learner with a constrained memory,
we may assume that there exists an abstract $\varepsilon$-machine 
in a larger space, here we consider a top-down approach: 
we start from stochastic processes that can be expressed as an
$\varepsilon$-machine, and 
apply quantum or classical compression methods to reduce the memory, at the cost 
of losing the accuracy in future predictions.
%In the analogy with the previous example with real-world data, the ``compressed model'' is the one available to a learner, with limited memory and imperfect accuracy, while the exact $\varepsilon$-machine is the abstract mathematical description given infinite resources. 
Since compression typically destroys the $\varepsilon$-machine structure, 
we lose the direct mapping between the classical and quantum simulators, 
which may then display  different accuracies. 
%Since compression impedes exact simulation, both the quantum and classical models can display different accuracy.
We introduce different compression methods, adapted from the tensor network literature, and discuss several figures of merit 
to investigate the trade-off between simulation accuracy and memory use, finding that the quantum 
advantage persists even in constrained memory situations. 

Finally, we consider whether the found quantum advantage in the top-down modelling results 
in a similar advantage in the bottom-up approach, which is closer to real-world scenarios. 
We train a model with a constrained memory via maximum likelihood and find that quantum models 
have higher accuracies.

\section{Notation and Background }
We consider systems described by random variables $X_t$, for $t\in\mathbb Z$,
which can take some discrete values $x_t\in\{1,\dots,d\}$ for some positive integer $d$. 
For clarity, we will refer to $t$ as a time index, so the $x_t$'s describe outcomes 
at different times, but the same formalism can model spatial correlations in 
one-dimensional systems \cite{feldman1998computational,crutchfield1997statistical}.
We assume that correlations between outcomes at different times (possibly far away) 
can be completely described via an auxiliary memory state, which 
is another time-dependent random variable $S_t$ with outcomes $s_t=\{1,\dots,D\}$ for 
some discrete memory dimension $D$. 

\begin{figure}[t]
	\centering
	\includegraphics[width=0.45\textwidth]{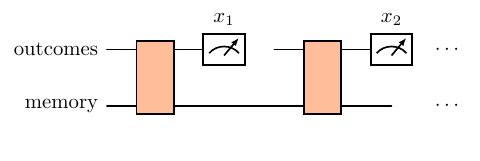}
	\caption{
		Pictorial representation of the evolution for the first two time steps. Both the classical and the quantum 
		simulators use two registers: an outcome register for the emitted symbols $x_t$, and a memory register 
		that keeps a compressed representation of the history of previous
		interactions and outcomes. At each time step, the 
		outcome register is first reset to a reference value, then it is let to interact with the memory 
		(orange box), and finally it is measured to observe the classical outcomes $x_t$. 
		For classical simulators, the orange box is mathematically modelled as a transition probability,
		Eq.~\eqref{eq:P def}, while for quantum simulators the orange box models a unitary operation,
		as in Eq.~\eqref{eq:U}, followed by a projective measurement to extract $x_t$.
	}
	\label{fig:cartoon}
\end{figure}

The model works as follows: starting from an initial memory state 
distribution $p^{\rm i}_\alpha = P(s_0{=}\alpha)$, the model emits the first outcome $x_1$ and 
then updates the internal memory state to a new value. This process is repeated 
at multiple times and an outcome sequence $x_1,x_2,\dots$ is generated. 
At a general time $t$, the model only knows the previous memory state and not
the entire history of outcomes and memory states. 
The transition probability from the previous memory state $s_{t-1}$ 
to the outcome $x_t$ and the new state $s_{t}$ is described by the transition 
probability tensor
\begin{equation}
	T^x_{\alpha,\beta} = P(x_t{=}x,s_t{=}\alpha|s_{t-1}{=}\beta),
	\label{eq:P def}
\end{equation}
where Latin and Greek indices repectively run from 1 to $d$ or from 1 to $D$. 
The above equation defines conditional probabilities, so in general $\sum_{x,\alpha} T^x_{\alpha,\beta}=1$ 
for each $\beta$. 
See Fig.~\ref{fig:cartoon} for a pictorial representation of the temporal process. 

The main assumption in Eq.~\eqref{eq:P def} 
is that $T$ is independent on $t$, namely that, at each time, knowledge of the memory state is sufficient 
to predict the outcomes. 
Models with these assumptions are called Hidden Markov Models (HMM), see e.g. \cite{murphy2012machine}, 
since the evolution of the memory state at time $t$ only depends on the memory state at time $t-1$
(and on the outcome $x_t$). In spite of the name, the distribution of the observed 
outcomes may be highly non Markovian, due to the hidden  memory states. 

Models described by \eqref{eq:P def} are 
called ``edge emitting'' HMM \cite{travers2011equivalence}, while it is customary for 
standard HMM to satisfy another property, namely that the state $s_t$ is independent on 
the outcome $x_t$. For such processes, $T$ factorises as 
$ T^{x}_{\alpha\beta} = J_{\alpha\beta} E^{x}_\beta $, namely as a product of 
the transition matrix 
$J_{\alpha\beta}=\sum_x T^x_{\alpha\beta}$ and an emission matrix 
$E^x_\beta=P(x_t{=}x|s_{t-1}{=}\beta)=\sum_\alpha T^x_{\alpha\beta}$. 
Such decomposition is particularly 
important in machine learning applications,
since the $E$ and $J$ matrices can be reconstructed from data using the 
Expectation-Maximization (EM) algorithm \cite{murphy2012machine}.
In this paper, we will consider the more general definition from Eq.~\eqref{eq:P def}. 

Ignoring the internal memory, the probability of observing the data sequence 
$\{x_1,\dots,x_L\}$ takes the matrix product form 
\begin{align}
	\label{eq:P mps explicit}
	P(x_1,\dots,x_L) &= \sum_{\{\alpha_t\}} T^{x_L}_{\alpha_L,\alpha_{L-1}}
	\cdots T^{x_1}_{\alpha_1,\alpha_{0}} p^{\rm i}_{\alpha_0} =\\ &= 
	\sum_{\alpha} (T^{x_L}\cdots T^{x_1} p^{\rm i})_\alpha,
	\label{eq:P mps}
\end{align}
where each $T^x$ is a matrix with indices as in Eq.~\eqref{eq:P def}, and $(v)_\alpha$ is the 
$\alpha$th component of vector $v$.

The initial memory 
state may be known from first principles or may be reconstructed from previous data. 
Suppose for instance that the data sequence can be split as an observed past $\{x_t\}_{t\leq 0}$
and an unknown future $x_1,\dots,x_L$. For a given transition tensor $T$ the initial memory 
probability can be reconstructed from the rules of conditional probability
\begin{equation}
	P({\rm future}|{\rm past})=\frac{P({\rm future, past})}{P({\rm past})}.
\end{equation}
Indeed, using the matrix product form \eqref{eq:P mps}, it is simple to realize 
that the exact future distribution given the past can be obtained  
by replacing $p^{\rm i}$ in Eq.~\eqref{eq:P mps} with an initial probability 
$p^{\rm i|past}$ that, up to a normalization factor, can be reconstructed as 
\begin{equation}
	p^{\rm i|past} \propto \left(\prod^{\leftarrow}_{x\in {\rm past}} T^x\right) p^{{\rm i}_0},
	\label{eq:conditional initial state} 
\end{equation}
where $p^{{\rm i}_0}$ is the initial probability
of the first observed event in the past. Without prior knowledge,
$p^{{\rm i}_0}$ can be considered uniform. 
The average state $	\mathbb E_{\rm past}\left[p^{\rm i|past}\right] $,
in the limit of many past observations, 
converges to the steady state $\pi$ of the transition matrix $J$,
\begin{align}
	\mathbb E_{\rm past}\left[p^{\rm i|past}\right] &\to \pi,  & 
	J\pi &= \pi,  &
	J = \sum_x T^x,
	\label{eq:steady classical}
\end{align}
namely the right-eigenvector of $J$ with largest eigenvalue -- for 
simplicity we assume that it is unique. 

%The transition tensor $T$ can be conveniently described as a state 
%transition matrix $J$ and an emission matrix $E^{x}_\alpha = P(x_t{=}x|s_{t-1}{=}\alpha)$ as 
%\begin{equation}
%	T^{x}_{\alpha\beta} = J_{\alpha\beta} E^{x}_\beta. 
%	\label{eq:EJ decomp}
%\end{equation}
%Note that from the definition $J_{\alpha\beta}=\sum_x T^x_{\alpha\beta}$ and 
%$E^x_\beta=\sum_\alpha T^x_{\alpha\beta}$. Such decomposition is particularly 
%important in machine learning applications where the transition tensor is unknown,
%since the $E$ and $J$ matrices can be reconstructed from data using the 
%Expectation-Maximization (EM) algorithm \cite{murphy2012machine}.
%However, such decomposition only holds when the final state depends 
%on the initial state only, and not on the outcome $x$. 

If we are interested in predicting future observations from a known past, we should 
keep the information from past observations that is relevant for predicting the future,
nothing more. 
In the asymptotic sense, the information about the past can be quantified 
by the Shannon entropy $H({\rm past}) = -\sum_\alpha \pi_\alpha \log_2\pi_\alpha$ 
of the steady state \eqref{eq:steady classical}, while the relevant information 
for predicting the future can be quantified by the mutual information 
$I({\rm future;past})$ or by the conditional entropy $H({\rm future}|{\rm past})=
H({\rm future}) - I({\rm future;past})$. The provably optimal model
\cite{crutchfield1989inferring,crutchfield1997statistical}
in this information theoretic sense, dubbed $\varepsilon$-machine, 
is the one that minimises $H({\rm past})$ 
so that $H({\rm future}|{\rm past}) = 0$, namely there is no uncertainty about 
the future given that the past is known. 
%so that the past has complete information to predict any arbitrary future. 
The memory compression in $\varepsilon$-machines is based on the intuition that we should not distinguish 
different pasts if they produce the same future. Mathematically this is performed by 
introducing an equivalence class $\sim$ such that 
two different past observations $x_{\rm past}$ and 
$x'_{\rm past}$  belong to the same equivalence class, namely $x_{\rm past}\sim x_{\rm past}'$,
if $P({\rm future}|x_{\rm past}) = P({\rm future}|x'_{\rm past})$ for arbitrary
future observations. 
The equivalence class is the only information we need to save in memory 
to predict the future. For complicated stochastic processes, 
the memory requirement may be infinite, though in this paper we only consider processes 
with finite memory. 
Numerically 
approximate $\varepsilon$-machine can be reconstructed from data using 
the Causal-State Splitting Reconstruction algorithm \cite{shalizi2002algorithm}.

The mapping $\varepsilon(x_{\rm past})$ from past observations to 
the corresponding equivalence class is deterministic. Hence, upon emitting 
a new observation $x_{t+1}$ the new memory state is deterministically 
obtained from the previous memory state and $x_{t+1}$. In other terms,
for $\varepsilon$-machines the transition probability \eqref{eq:P def}
takes the form 
\begin{equation}
	T^x_{\alpha\beta} =  \delta_{\alpha,\sharp(x, \beta)}E^x_\beta,
	\label{eq:unifilar}
\end{equation}
where $\sharp(x,\beta)$ is a function mapping the previous memory state $\beta$ 
and the emitted outcome $x$ into a new memory state, and 
$\delta$ is the Kronecker function. HMMs satisfying the above 
property are called unifilar \cite{travers2011equivalence}. 
%Unifilar Markov processes are important in the context of $\varepsilon$-machines, namely HMMs where the hidden states are equivalence classes of past observations that produces the same future. Two different past observations $x_{\rm past}$ and $x'_{\rm past}$  belong to the same equivalence class if $P({\rm future}|x_{\rm past}) = P({\rm future}|x'_{\rm past})$ for arbitrary future observations. Such equivalence classes and the resulting approximate $\varepsilon$-machine can be reconstructed from data using the Causal-State Splitting Reconstruction algorithm \cite{shalizi2002algorithm}.

\subsection{Quantum Simulation}\label{s:quantum simulators} 
Quantum simulators of HMMs have been proposed, both for unifilar
\cite{gu2012quantum,yang2018matrix} and not-unifilar 
\cite{elliott2021memory} processes. In both cases, the mapping exploits a
quantum memory register,  indexed by some normalized but non-orthogonal vectors
$\{\ket{\sigma_\alpha}\}$, in place of the classical memory,
where $\alpha=1,\dots,D$, as before. 
In the unifilar case, the unitary simulator of an HMM with transition tensor $T$ is defined as 
\begin{equation}
U\ket{0,\sigma_\beta} = \sum_{x\alpha} \sqrt{T^x_{\alpha\beta}} \ket{x,\sigma_\alpha},
\label{eq:U}
\end{equation}
where the outcome register $\ket{x}$ is initialized in a reference state (say
$\ket{0}$) -- see also Fig.~\ref{fig:cartoon}. 
By iteratively applying $L$ times such operator $U$, each time measuring the reference state and  reinitializing 
it in $\ket0$, %and applying $U$, 
we get a state that acts on $L$ tensor copies of the outcome Hilbert space,
and a single memory space. More precisely we can expand 
\begin{equation}
	\ket{\psi[A]} = \sum_{\{x_t\}} \sqrt{P_A(x_{1:L})} \ket{x_{1:L}}\ket{\psi_S(x_{1:L})} ,
	\label{eq:psi expansion}
\end{equation}
where we have defined $x_{1:L}=(x_1,\dots,x_L)$, 
$A^x_{\alpha\beta}= \sqrt{T^x_{\alpha\beta}} $ and
\begin{align}
	\ket{\psi_S(x_1,\dots,x_L)} &= \frac{A^{x_L}\cdots A^{x_1}\ket{\sigma}}{\sqrt{P_A(x_1,\dots,x_L)}}
	\label{eq:quantum from A}
	\\
%	A^x\ket{\sigma_\beta} & = \sum_\alpha \sqrt{T^x_{\alpha\beta}} \ket{\sigma_\alpha}, \\ 
	P_A(x_1,\dots,x_L) &= \|A^{x_L}\cdots A^{x_1}\ket{\sigma}\|_\sigma^2.
	\label{eq:P unifilar}
\end{align}
In the above equations we have assumed 
an initial memory state $\ket{\sigma}$ and
introduced the norm in the non-orthogonal basis
$\|\ket{\psi}\|_\sigma^2 = \sum_\alpha |\psi_\alpha|^2$ 
for a generic state $\ket{\psi} = \sum_\alpha \psi_\alpha
\ket{\sigma_\alpha}$. 

For unifilar processes, after tracing out the memory index, the probability that we get from 
Eq.~\eqref{eq:P unifilar} is equal to the probability of the data sequence \eqref{eq:P mps},
in other terms $P(x_1,\dots,x_L)=P_A(x_1,\dots,x_L)$.
This is a consequence of Eq.~\eqref{eq:unifilar}, namely that, for a given input memory state and emitted outcome, 
the next memory state is deterministic. Accordingly, the sums over the memory
indices in Eqs.~\eqref{eq:P mps explicit} 
and~\eqref{eq:quantum from A} are removed, as there is only a possible string of
$\{\alpha_t\}$, and after taking the square in Eq.~\eqref{eq:P unifilar},
for a suitable input state, we recover exactly the expression of Eq.~\eqref{eq:P mps explicit}. 

In order to define the mathematical properties of quantum simulators it is more convenient 
to normalize the tensors in Eq.~\eqref{eq:quantum from A} so that they define 
normalized Kraus operators \cite{nielsen2002quantum}. Let 
$G_{\alpha\beta}=\bra{\sigma_\alpha}\sigma_\beta\rangle$ be the Gram matrix of the memory states.
From the unitarity of the operator $U$ in Eq.~\eqref{eq:U} we get 
$G_{\alpha\beta}=\bra{0,\sigma_\alpha}U^\dagger U\ket{0,\sigma_\beta}$, which 
leads to the following equation 
\begin{equation}
	G = \sum_x A^x{}^\dagger G A^x,
	\label{eq:Gram steady}
\end{equation}
namely that $G$ is the fixed point of the map $\mathcal E^\dagger_A(\cdot) = 
\sum_x A^x{}^\dagger (\cdot) A^x$. Since Gram matrices are positive, we may decompose them
as $G=W^\dagger W$, where $W_{\alpha\beta}=\bra{\alpha}\sigma_\beta\rangle$ 
and $\ket{\alpha}$ are orthogonal vectors. Accordingly, the operators 
\begin{equation}
	K^x = W A^x W^{-1},
	\label{eq:Krauss}
\end{equation}
define Krauss operators of a completely positive map, since from Eq.~\eqref{eq:Gram steady} 
we get $\sum_x K^x{}^\dagger K^x=\openone$. Moreover, since in matrix product states 
the actions of $W$ and $W^{-1}$ cancel out, the probability of outcomes is the same. 
More precisely, using $W$ we may write the norm in Eq.~\eqref{eq:P unifilar} as 
$\| \ket\psi \|_\sigma^2 = \| W \ket{\psi}\|_2^2 = \bra\psi G \ket{\psi}$. Accordingly,
setting $\ket{\phi}=W^{-1}\ket{\sigma}/
\|W^{-1}\ket{\sigma}\|$ and defining 
\begin{equation}
	P_K(x_1,\dots,x_L) = \|K^{x_L}\dots K^{x_1}\ket{\phi}\|,
	\label{eq:prob K}
\end{equation}
%calling $P_K(x_1,\dots,x_L)$ the probability \eqref{eq:P unifilar} when all 
%$A^{x_t}$ are replaced by $K^{x_t}$, 
then we get $P_K(x_1,\dots,x_L)= P_A(x_1,\dots,x_L)$.
The above evolution can be generally modelled as in Fig.~\ref{fig:cartoon} with unitary 
interactions between the memory and outcome registers. 
A more general case is discussed in Appendix~\ref{app:nonorm}.

We now focus on the memory state, given that we have obtained some observations 
$\{x_t\}$. From Eq.~\eqref{eq:psi expansion}, given a trial initial state $\ket\phi$
we see that this is given by 
\begin{equation}
	\frac{
		K^{x_L}\cdots K^{x_1}\ket{\phi}\bra\phi K^{x_1}{}^\dagger \cdots K^{x_L}{}^\dagger 
	}{P_K(x_1,\dots,x_L)},
\end{equation}
appearing with probability $P_K(x_1,\dots,x_L)$.
The average memory state after $L$ observations is then 
\begin{align}
	\rho_L &= \sum_{\{x_t\}} 
	K^{x_L}\cdots K^{x_1}\ket{\phi}\bra\phi K^{x_1}{}^\dagger \cdots K^{x_L}{}^\dagger 
			\\ &
			= \mathcal E_K^L(\ket{\phi}\bra\phi) \stackrel{L\to\infty}{\longrightarrow} \Pi ,
\end{align}
%which in the limit of large $L$ converges to the steady state $\Pi$ of the map 
where $\Pi$ is the fixed point (steady state) of the map 
$\mathcal E_K(\cdot) = \sum_x K^x (\cdot) K^x{}^\dagger$. This state can be obtained 
from the steady state \eqref{eq:steady classical} of the classical model by noting that, from
the definition of $A$ and Eq.~\eqref{eq:steady classical}, we get 
$\sum_x A^x {\rm diag}(\pi) A^x{}^\dagger = {\rm diag}(\pi)$. This may in turn 
be rewritten as 
\begin{equation}
	\sum_x K^x \Pi K^x{}^\dagger = \Pi,
	\label{eq:krau steady}
\end{equation}
so the steady state of the map $\mathcal E_K$ takes the form
\begin{equation}
	\Pi = \sum_\alpha \pi_\alpha \ket{\sigma_\alpha}\!\bra{\sigma_\alpha}
	= W {\rm diag}(\pi)W^\dagger.
	\label{eq:Pi pi}
\end{equation}
Note that, since $W$ is not a unitary matrix, 
the classical steady state does not define the eigenvalues of the quantum steady state 
in \eqref{eq:Pi pi}. This is the main difference responsible for the memory advantage 
in the quantum simulation of classical stochastic processes \cite{gu2012quantum,yang2018matrix}. 
Indeed, since the entropy of a mixture of pure quantum states is smaller than 
the entropy of the mixture, 
\begin{equation}
	H(\Pi) \leq H(\pi)
	\label{eq:memory advantage}
\end{equation}
where $H(\Pi) = -\Tr[\Pi\log_2\Pi]$ is the Von Neumann entropy of the steady state. Equality 
in \eqref{eq:memory advantage} arises only when $\ket{\sigma_\alpha}\!\bra{\sigma_\alpha}$
are orthonormal \cite{nielsen2002quantum}. 
When this is not the case, quantum simulators 
can use significantly less memory to model the same classical stochastic process with exact 
accuracy \cite{elliott2021memory,yang2018matrix}.
The key intuition is that  $\varepsilon$-machines define in many cases an 
irreversible process and, in many standard examples, $H(\pi)$ is 
strictly greater than $I({\rm past};{\rm future})$, which quantifies the ultimate amount
of needed information. Improvements in the 
quantum simulation arise from addressing the source of irreversibility within 
quantum dynamics \cite{gu2012quantum}.

Beyond the unifilar case, unitary simulation of an HMM with transition tensor $T$ is 
possible by using matrix product density operators, as shown in Appendix \ref{a:not unifilar}.
In this work we only consider quantum simulators based on matrix product states, where 
the formal mapping between classical and quantum simulators is valid for 
unifilar HMM only. Nonetheless, we will assume a structure like in Eq.~\eqref{eq:quantum from A} and 
study the accuracy of the resulting simulator for general (possibly not-unifilar) stochastic 
processes.

\section{Memory Compression}\label{sec:compress}

%The amount of randomness in the memory state before reading it, or in other terms the amount of information we get about the memory after reading, can be quantified by entropy. Specifically, for classical systems \cite{feldman1998computational} one focuses on the equilibrium state \eqref{eq:steady classical}, while in quantum systems one may focus on the equilibrium density matrix \eqref{eq:Pi pi}. Calling $H(\pi) = -\sum_\alpha \pi_\alpha \log_2\pi_\alpha$ the Shannon entropy and $H(\rho) = -\Tr[\rho\log_2\rho]$ the Von Neumann entropy, it has been shown that quantum simulators can use significantly less memory to model the same classical stochastic processes \cite{elliott2020extreme,elliott2021memory,yang2018matrix}, with exact accuracy.

Here we study whether the memory advantage of quantum simulators
persists even when we relax the requirement of 
exact simulation. We introduce different compression methods for either classical or 
quantum simulators and introduce different measures of accuracy.

\subsection{Memory compression of quantum simulators}

Compressions methods for matrix product states were discussed in 
several papers, see e.g.~\cite{orus2008infinite,schollwock2011density,vidal2007entanglement}.
Here we focus on adapting the spectral compression method from \cite{orus2008infinite}
in order to maintain a high overlap between the original and compressed states, which is 
expected to provide a large fidelity and, accordingly, an accurate prediction of future 
observations.

\begin{figure}[t]
\begin{algorithm}[H]
	\caption{Compress the quantum simulator $\ket{\psi[A]}$}
  \label{alg:quantum compress}
   \begin{algorithmic}[1]
		 \Require{ A normalized $d\times D\times D$ tensor $A$ and the dimension $D'<D$ of the compressed memory}
		 \Function{mpscompress}{$A,D'$}
		 \State Compute the steady state $\Pi_A$ as the eigenoperator 
		 \StatexIndent[1.5] with largest 
		 eigenvalue $1$ of the map 
		 \StatexIndent[1.5] $\mathcal E_A[\rho]= \sum_x A^x\rho A^{x\dagger}$.
		 \State Find the spectral decomposition of $\Pi_A = W\lambda W^\dagger$, 
		 \StatexIndent[1.5] where 
		 $W$ is unitary and $\lambda$ is diagonal with diagonal \StatexIndent[1.5] 
		 elements sorted in decreasing order.
		 \State Define the tensor $B$ with elements 
		 \StatexIndent[1.5] $B^x = P' W^\dagger A^x W P'$, where $P'$ is a
		 a $D'\times D$ 
		 \StatexIndent[1.5] projection operator with non-zero elements $P'_{ii}=1$. 
		 \State \Return \Call{mpsnormalize}{$B$} 
		 \EndFunction
   \end{algorithmic}
\end{algorithm}
\vspace{-5mm}
\begin{algorithm}[H]
	\caption{Normalize the quantum simulator $\ket{\psi[A]}$}
  \label{alg:normalize}
   \begin{algorithmic}[1]
		 \Require{ A $d\times D\times D$ tensor $A$}
		 \Function{mpsnormalize}{$A$}
		 \State Compute $G$ as the eigenoperator of the map 
		 \StatexIndent[1.5] $\mathcal E^\dagger_A[Y]= \sum_x A^{x\dagger}Y A^{x}$ whose corresponding 
		 \StatexIndent[1.5] eigenvalue $\mu$ has largest $|\mu|$.
		 \State Find the spectral decomposition of $G = UsU^\dagger$, where 
		 \StatexIndent[1.5] $s$ is diagonal and positive semi-definite.
		 \State \Return the tensor $B$ with elements 
		 \StatexIndent[1.5] $B^x = s^{1/2} U^\dagger A^x U s^{-1/2}/\sqrt\mu$
		 \EndFunction
   \end{algorithmic}
\end{algorithm}
%\vspace{-5mm}
%	\caption{Normalization and compression algorithms for quantum simulators.}
%	\label{fig:algos12}
\end{figure}

Suppose that we have an exact quantum simulator $\ket{\psi[A]}$, where the tensor $A$ has 
memory dimension $D$. We want to find a new simulator $\ket{\psi[B]}$ where the $B$ 
tensor has reduced dimensionality $D'<D$, such that the similarity of future observations 
is as high as possible
and all the properties of the original simulator are approximately maintained. 
%We first note that, ignoring the memory index in \eqref{eq:psi expansion}, we get 
%a mixed state $\rho[A]=\Tr_{\rm mem}[\ket{\psi[A]}\bra{\psi[A]}]$, whose diagonal 
%elements are exactly $P_A(x_{1:L})$. 
%$F(\rho[A],\rho[B])$ is as high as possible 
For instance, calling $\Pi_Y = \mathcal E_Y[\Pi_Y]$, for $Y$ either $A$ or $B$, the steady state of 
$\mathcal E_Y[\cdot] = \sum_x Y^x (\cdot) Y^{x\dagger}$, we want the asymptotic 
states of the two simulators to share a similar amount of information, namely $H(\Pi_A) \simeq H(\Pi_B)$. 
With these goals in mind, we define Algorithm~\ref{alg:quantum compress}. 
%, shown in Fig.~\ref{fig:algos12}. 
The main ideas behind such algorithm are as follows: we first transform 
the tensors $A$ with a change of basis, in such a way that the equilibrium memory state $\Pi_A$ 
is diagonal, with diagonal elements $\lambda_i$ sorted in decreasing order. In this basis
we then truncate the memory indices, selecting the first $D'$ elements. The obtained tensors are then 
normalized using Algorithm~\ref{alg:normalize}. If the truncated elements 
$\lambda_i$ are small and the normalization operation does not alter too much the tensors
(which can be expected from perturbation theory \cite{kato2013perturbation}), then 
the steady state of the resulting tensor is expected to have a similar memory entropy.

\subsection{Memory compression of classical simulators}

\begin{figure}[t]
\begin{algorithm}[H]
	\caption{Compress a Hidden Markov Model} 
  \label{alg:classical compress}
   \begin{algorithmic}[1]
		 \Require{ A normalized $d\times D\times D$ transition tensor  $T$ and the
		 dimension $D'<D$ of the compressed memory}
		 \Function{hmmcompress}{$T,D'$}
		 \State Compute the transition matrix $J_{\alpha\beta}= \sum_x T^{x}_{\alpha\beta}$, 
		 \StatexIndent[1.5] 
		 the steady state $J\pi=\pi$ and the recovering and 
		 \StatexIndent[1.5] 
		 compression matrices from Eqs.~\eqref{eq:reduced steady}-\eqref{eq:classical encode}.
		 \State Set $ T'^x_{\alpha'\beta'}= \sum_{\alpha,\beta=1}^D C_{\alpha',\alpha}T^x_{\alpha\beta} R_{\beta,\beta'}$ for
		 \StatexIndent[1.5] 
		 $\alpha',\beta'=1,\dots,D'$
		 \State \Return $T'$
		 \EndFunction
   \end{algorithmic}
\end{algorithm}
\end{figure}

Different methods have been proposed to reduce the memory requirements of 
classical Markov chains \cite{geiger2022information}. 
Inspired by Algorithm~\ref{alg:quantum compress}, we develop two memory compression 
methods for classical HMMs. The first one is based on an encoding/decoding protocol 
aimed at preserving the entropy of the steady state, while the second one, 
presented in Appendix~\ref{a:cl compress} is based on spectral compresion methods 
adapted from \cite{zhang2019spectral,wu2010probability}. 

Consider the steady state of the transition matrix defined in Eq.~\eqref{eq:steady classical}. 
As described in the previous section, Algorithm~\ref{alg:quantum compress} has the desired 
properties of compressing the state space while keeping as much information as possible 
about the asymptotic states. We can develop a classical compression method with a similar 
property as follows. Let $D'<D$ be the reduced dimension and assume that $\pi_\alpha$ 
are in decreasing order. Setting 
\begin{align}
	\pi'_{\alpha'} &= \pi_\alpha / \lambda & \lambda = \sum_{\alpha' = 1}^{D'} \pi_{\alpha'}
	\label{eq:reduced steady}
\end{align}
for $\alpha'=1,\dots, D'$ as the compressed steady state, we aim at designing a coding 
and decoding strategy so that $\pi'$ is the steady state of the compressed HMM. 
We can define a recovering protocol as a $D\times D'$ transition matrix
$R_{\alpha,\alpha'} = P(\alpha|\alpha')$ 
so that $R\pi' = \pi$ as follows 
\begin{equation}
	R = \begin{pmatrix}
		\lambda &&&\\
		&\lambda &&\\
		&&\ddots&\\
		&&&\lambda\\
		\pi_{D'+1} &&\cdots & \pi_{D'+1} \\
		\vdots && & \vdots \\
		\pi_{D} &&\cdots & \pi_{D} \\
	\end{pmatrix}.
	\label{eq:classical decode}
\end{equation}
Notice that, by construction, $R$ defines a conditional distribution and $\sum_\alpha R_{\alpha,\alpha'} = 1$. 
The encoding protocol can now be constructed according to the Bayes rule
\begin{equation}
	C_{\alpha',\alpha} = P(\alpha'|\alpha) = \frac{P(\alpha|\alpha') P(\alpha')}{P(\alpha)} 
= \frac{R_{\alpha,\alpha'} \pi'_{\alpha'}}{\pi_\alpha}.
	\label{eq:classical encode}
\end{equation}
The resulting procedure is summarized in Algorithm~\ref{alg:classical compress}.

\subsection{Quantifying the accuracy of a compressed quantum simulator} 
Suppose we have two quantum simulators $\ket{\psi[A]}$ and $\ket{\psi[B]}$, normalized 
in such a way that $A^x$ and $B^x$ define two different sets of Kraus operators. 
In what follows we assume that $\ket{\psi[A]}$ provides an exact simulation of 
the process, while $\ket{\psi[B]}$ is an approximation, using a reduced memory.
The statistical distance (e.g.~the Kullback-Leibler or Bhattacharyya distance) 
between the two future outcomes distributions,
$P_A(x_{1:L})$ and
$P_B(x_{1:L})$, can be computed explicitly for a given reasonably small 
$L$, but its numerical complexity grows exponentially with $L$. 
Because of this, we introduce a different figure of merit based on the fidelity between 
quantum states, which can be computed efficiently for matrix product operators \cite{hauru2018uhlmann},
allowing us to study even the limit $L\to\infty$. Similar measures have been previously 
considered in \cite{yang2020measures}.

We first note that, ignoring the memory index in \eqref{eq:psi expansion}, we get 
a mixed state $\rho[A]=\Tr_{\rm mem}[\ket{\psi[A]}\bra{\psi[A]}]$, whose diagonal 
elements are exactly $P_A(x_{1:L})$. 
From the data processing inequality, $F(\rho[A],\rho[B])$ is smaller than 
the Bhattacharyya distance between $P_A$ and $P_B$. 
Using the Uhlmann's theorem 
and the matrix product state expansion we may write 
\begin{align}
	F(\rho[A],\rho[B]) &= \max_{U : U U^\dagger=\openone}  |\bra{\psi[B]}U_{\rm memory}\ket{\psi[A],0_p}|
									\\ &
	= \max_{X : \|X\|_\infty \leq 1} |\bra{\psi[B]}X_{\rm memory}\ket{\psi[A]}| \\&
	= \max_{X : \|X\|_\infty \leq 1} |\Tr[X \mathcal E_{A,B}^L[\ket{\sigma_A}\bra{\sigma_B}]]|\\&
= \|\mathcal E_{A,B}^L[\ket{\sigma_A}\bra{\sigma_B}]\|_1,
\label{eq:fidelity quantum}
\end{align}
where $U$ is a unitary matrix acting on the memory space, $\ket{0_p}$ is a ``padding'' ancillary 
state to make the dimension of the memory registers equal, % (assuming that $B$ has reduced dimensionality),
$\|X\|_p = \sqrt[p]{\Tr(X^\dagger X)^p}$ is the Schatten $p$-norm, $\ket{\sigma_{A/B}}$ are the initial memory 
memory states of $\ket{\psi[A/B]}$ and we have defined the linear map
\begin{equation}
	\mathcal E_{A,B}[Y] = \sum_x A^x Y B^{x\dagger}.
	\label{eq:fidelity map}
\end{equation}
Note that, in general, $\mathcal E_{A,B}$ maps rectangular operators into rectangular operators, 
since the $A^x$ and $B^x$ may have different dimensions. 
For large $L$ the operator power $\mathcal E_{A,B}^L[Y]$ can be approximated as 
$\mathcal E_{A,B}^L[Y] \approx \lambda_{A,B}^L \Lambda_R \Tr[\Lambda_L Y]$ 
where $\lambda_{A,B}$ is the eigenvalue of $\mathcal E_{A,B}$ with largest absolute value,
and $\Lambda_{L/R}$ are the corresponding left/right eigenoperators. Assuming that $\lambda_{A,B}$ is 
unique we then get $F(\rho[A],\rho[B])\propto |\lambda_{A,B}|^L$. This derivation
also provides a way of selecting the compressed memory state $\ket{\sigma_B}$ given 
the initial state $\ket{\sigma_A}$. Indeed, it appears from Eq.~\eqref{eq:fidelity quantum}
that the optimal state is that maximising $|\bra{\sigma_B} \Lambda_L \ket{\sigma_A}|$, namely 
up to a normalization $\ket{\sigma_B} \propto \Lambda_L \ket{\sigma_A}$.

From the fidelity we may define the R\'enyi divergence between two density matrices \cite{muller2013quantum},
as $D_{1/2}(A\| B) = -2 \log F(\rho[A],\rho[B])$. Our first measure of accuracy is then the 
divergence density (see also \cite{yang2020measures})
\begin{equation}
	H(A\|B) = \lim_{L\to\infty} \frac{D_{1/2}(A\|B)}{2L} = -\log|\lambda_{A,B}|,
	\label{eq:divergence q}
\end{equation}
where, as before, $\lambda_{A,B}$ is the eigenvalue of the map \eqref{eq:fidelity map} with 
largest absolute value. 

The above divergence measures the asymptotic decay rate of the fidelity, without considering 
the effect of the initial memory state. To consider the effect of past observations without 
making any assumptions on the initial guess and observed data, 
we focus on the average asymptotic behaviour described by the steady 
state \eqref{eq:Pi pi}. Calling $\ket{\Pi_{A/B}}$ the purification of the steady states 
of maps $\mathcal E_{A/B}$ and extending the derivation of Eq.~\eqref{eq:fidelity quantum} we get 
\begin{equation}
	F(\rho_\Pi[A],\rho_\Pi[B]) =
\|(\mathcal E_{A,B}^L \otimes \mathcal I)[\ket{\Pi_A}\bra{\Pi_B}]\|_1,
\label{eq:fidelity steady}
\end{equation}
where $\mathcal I$ is an identity channel, and $\rho_\Pi$ refers to a simulator with initial memory state
described by the steady mixed state \eqref{eq:krau steady}.

\subsection{Quantifying the accuracy of a compressed classical simulator}\label{sec:compress cl} 

For quantum simulators we use the fidelity $F(\rho[A],\rho[B])$ to quantify the accuracy between 
the exact and compress quantum simulators. The closest classical measure is the Bhattacharyya 
coefficient $B[T,\bar T] = \sum_{\{x_t\}} \sqrt{P(x_{1:L})\bar P(x_{1:L})}$, where $\bar P$ 
is alike $P$ in Eq.~\eqref{eq:P mps explicit} but with truncated transition tensor $\bar T$. 
However, $B[T,\bar T]$ is difficult to compute for larger $L$ since we cannot make explicit 
use of the matrix product form \eqref{eq:P mps} for avoiding to sum over the $d^L$ outcomes. 
Other measures \cite{cha2007comprehensive} based on the evaluation of 
the inner product $P\cdot \bar P= 
\sum_{\{x_t\}} P(x_{1:L})\bar P(x_{1:L})$ can be explicitly computed via 
matrix powers of the linear map $\mathcal E_T[\cdot] =
\sum_x T^x(\cdot)\bar T^{xT}$. The latter can be done efficiently in a time the scales linearly in $L$,
rather than exponentially. Examples of such similarity measures include the 
cosine similarity $P\cdot \bar P/\sqrt{(P\cdot P)(\bar P\cdot \bar P)}$, which was also considered 
in \cite{yang2020measures}.

\section{Numerical results} 
We focus on stochastic processes with an exact representation as an 
$\varepsilon$-machines with finite memory, for which the quantum advantage in the exact 
simulation can be formally assessed \cite{gu2012quantum}, and we study whether such an advantage 
persists when we relax the assumption of exact simulation, by reducing the available memory. 

We consider the $\varepsilon$-machine of discrete renewal processes 
\cite{marzen2017informational} of period $N$, a kind of stochastic clock 
where the number of 0s between two consecutive ``ticks'' (with outcome 1) 
is uniformly distributed between 0 and $N-1$. As $N$ increases the process becomes 
increasingly more non-Markovian as we need to store the last $N$ entries to predict 
future statistics. It is clear though that this can be represented as a HMM with 
memory $N$, since we can simply store the number of 0s since the last tick. 

The transfer matrix 
of this process is given by 
\begin{align}
	%T^0 &= \sum_{k=1}^{N-1}\frac{(N-k)\ket{k+1}\bra k}{N+1-k}, &
	T^0_{k+1,k} &= \frac{N-k}{N+1-k}, &
	T^1_{1,k} &= \frac1{N+1-k},
	%T^1 &= \sum_{k=1}^N\frac{\ket 1\bra k}{N+1-k},
\end{align}
where $k=1,\dots,N$. The normalized Kraus operators of the quantum simulator have 
the analytic form \cite{yang2018matrix}
\begin{align}
	K^0 &= \sum_{k=1}^{N-1}\ket{k+1}\bra k, &
	K^1 &= \sum_{k=1}^N \frac{\ket{k}\bra N}{\sqrt N}.
\end{align}

\begin{figure}[t]
	\centering
	\includegraphics[width=0.42\textwidth]{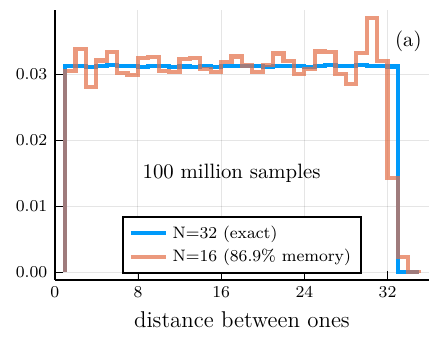}
	\includegraphics[width=0.42\textwidth]{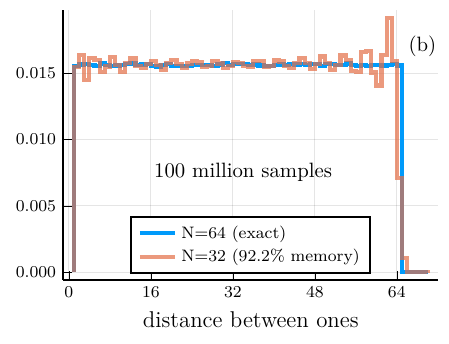}
	\caption{Distribution of the distance between consecutive ones in the exact and 
	compressed quantum simulation of of a discrete renewal process of period $N$, 
	for $N=32$ (a) or $N=64$ (b).
	}
\label{fig:cmpsamples}
\end{figure}

In Fig.~\ref{fig:cmpsamples} we compare the performance of the exact and compressed 
quantum simulators, obtained using Algorithm~\ref{alg:quantum compress}. We initialize 
both simulators in the memory states $\sum_i \sqrt{\lambda_i}\ket{\lambda_i}$ where $\lambda_i$ 
and $\ket{\lambda_i}$ are, respectively, the eigenvalues and eigenvectors of their 
steady state \eqref{eq:krau steady} and, for each simulator, we compute $10^8$ samples. 
Getting so many samples is easy, because the probability distribution \eqref{eq:prob K} 
can be factorized. The full procedure  is illustrated in Algorithm~\ref{alg:sample}.

\begin{figure}[t]
\begin{algorithm}[H]
\caption{Sample from the quantum simulator}
  \label{alg:sample}
   \begin{algorithmic}[1]
		 \Require{ A normalized $d\times D\times D$ tensor $A$, 
		 an initial memory state $\ket{\phi}$, and the number of observations $L$.}
		 \Function{mpssample}{$A,\ket{\phi},L$}
		 \For{$t=1,\dots,L$}
		 \State Sample $x_t$ from the probability distribution 
		 \StatexIndent[2.5] $P(x)=\|A^x\ket{\phi}\|^2$.
		 \State Update $\ket{\phi}$ to $A^{x_t}\ket{\phi}/\sqrt{P(x_t)}$.
		 \EndFor
		 \State \Return observations $x_1,\dots,x_L$.
		 \EndFunction
   \end{algorithmic}
\end{algorithm}
%\caption{}
%\label{fig:sample}
\end{figure}

As shown in Fig.~\ref{fig:cmpsamples}, the exact simulator reproduces almost perfectly 
the expected flat distribution in the distance between ones, with the differences due 
to the finite number of samples. The compressed quantum simulators, where the memory 
dimension has been cut to half the original one, still captures the main features 
of the distribution, though with some fluctuations. As expected for
the chosen stochastic process, errors tend to be larger when the distance between ones 
is close to the maximum value $N$. We note in particular that the distribution 
oscillates around the expected value, and that the amplitude of the oscillations increases
for larger distances between ones. Around the maximum distance $N$, there is a significant drop with 
a large deviation from the expected probability. 

The entropy of the exact and compressed simulators are respectively 1.26 and 1.16 for $N=64$, 
1.23 and 1.07 for $N=32$. Therefore, although the compressed simulators have a halved 
memory space, the entropy of the steady states are reduced by just $8{\div}13\%$. 
This shows that the spectral compression method introduced in Algorithm \ref{alg:quantum compress} 
achieves the desired goal of reducing the number of memory states, while keeping a similar amount 
of entropy and maintaining the ability to predict future outcomes. 
To put things into perspective, we also compute the entropy of the truncated steady 
states. Namely keeping half of the largest eigenvalues of the steady states, normalizing them 
and computing the entropy we get 1.19 for $N=64$ and 1.13 for $N=32$. The further reduction 
provided by Algorithm~\ref{alg:quantum compress} is therefore given to the normalization 
procedure of Algorithm~\ref{alg:normalize}, that reduces the entropy by a
further $5\%$ compared to the optimal case. 

\begin{figure}[t]
	\centering
	\includegraphics[width=0.42\textwidth]{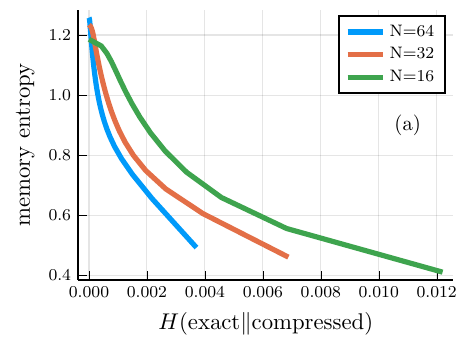}
	\includegraphics[width=0.42\textwidth]{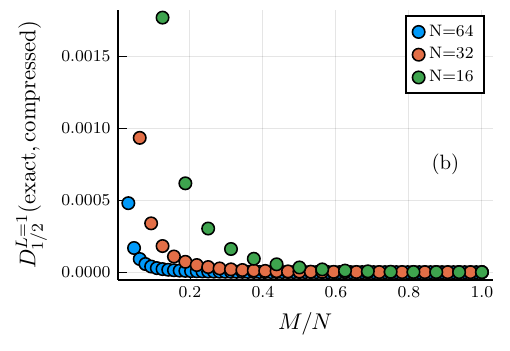}
	\includegraphics[width=0.42\textwidth]{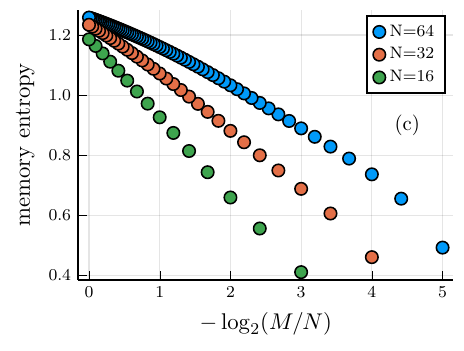}
	\caption{Memory entropy vs accuracy  between the exact 
	and compressed simulators. For each $N$, the compressed simulators are 
	computed with truncated memory dimension $M=2,\dots,N$. 
	%When $M=N$ the memory is maximal and $D=0$. The memory entropy decreasing monotonically for decreasing $M$, while $D$ increases monotonically. The points with larger $D$ are for $M=2$. 
	(a) Entropy of compressed memory vs Eq.~\eqref{eq:divergence q}. 
	(b) Divergence $D_{1/2}$ for $L=1$ vs $M/N$. 
	(c) Entropy vs $-\log_2(M/N)$.
	}
	\label{fig:fcast}
\end{figure}

In Fig.~\ref{fig:fcast} we study how entropy and accuracy, as quantified by the 
fidelity \eqref{eq:fidelity quantum} or by its asymptotic 
decay rate \eqref{eq:divergence q}, change as a function of the truncated memory. 
In Fig.~\ref{fig:fcast}(b) we see that the divergence $D_{1/2} = -2\log F$, where
$F$ is the fidelity between exact and truncated simulators, is negligible, so 
we can focus on the decay rate shown in Fig.~\ref{fig:fcast}(a). 
For instance, with a decay rate $H=0.004$ the fidelity behaves as 
$F\approx e^{-0.004L}$ and reaches $F\approx 50\%$ for $L\simeq 170$. 
Therefore, we need to make predictions far ahead in the future to observe 
a significant reduction in accuracy. 
In Fig.~\ref{fig:fcast}(c) we study the decay in memory for 
increasing $N$, and observe an almost linear decrease as a function of $-\log_2(M/N)$.
When $M=N$ the simulation is exact, while for $M=2$ (rightmost points for every color)
we use the minimum amount of memory. 
From the monotonic behaviour of Figs.~\ref{fig:fcast}(b) and (c), we understand that 
the entropy vs accuracy trade-off curve shown in Fig.~\ref{fig:fcast}(a) is explored 
from left to right by reducing $M$ from $N$ to 2. This curve shows the performances 
of the compression Algorithm~\ref{alg:quantum compress} in keeping the accuracy 
of quantum simulators while reducing the memory cost.

\begin{figure}[t]
	\centering
	\includegraphics[width=0.42\textwidth]{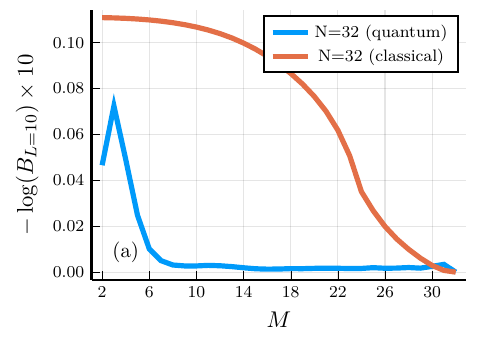}
	\includegraphics[width=0.42\textwidth]{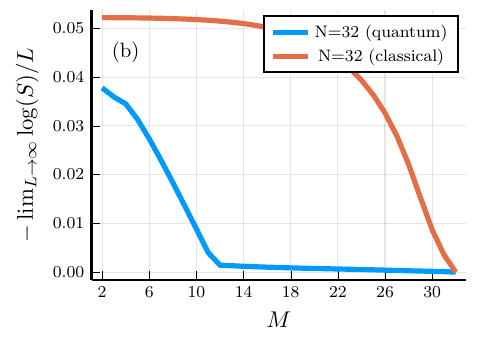}
	\includegraphics[width=0.42\textwidth]{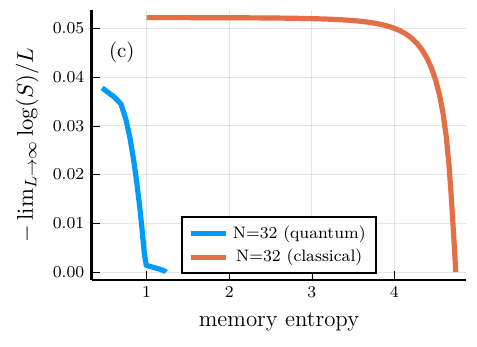}
	\caption{
		Compressed memory dimension $M$ vs.~Bhattacharyya coefficient (a) 
		or vs.~the similarity decay rate for $L\to\infty$ (b)
		in the simulation of a discrete renewal process with $N=32$ and $M\leq N$. 
		(c) Memory entropy vs.~similarity decay rate in the same setting of (a-b). 
		%Compressed memory dimension $M$ vs.~similarity decay rate for $L\to\infty$ (a) and 
		%vs.~entropy (b) in the simulation of a discrete renewal process with $N=32$ and $M\leq N$. 
	}
	\label{fig:qvsc32}
\end{figure}

Finally, in Fig.~\ref{fig:qvsc32} we compare the performance of Algorithms~\ref{alg:quantum compress} 
and \ref{alg:classical compress} in compressing, respectively, the exact classical and quantum simulators 
of discrete renewal stochastic processes with $N=32$. Since for classical simulators we cannot compute the fidelity, 
we focus on the Bhattacharyya coefficient $B$ and 
the similarity measure $S$ introduced in Sec.~\ref{sec:compress cl}. The Bhattacharyya coefficient 
is computed by first generating a dataset of 1000 samples using Algorithm~\ref{alg:sample}, 
setting the memory according 
to the Bayes rule Eq.~\eqref{eq:conditional initial state}, and then predicting the probability of 
future observations $x_{1:L}$ for $L=10$. 
On the other hand, the similarity is expected to decay as $S\propto e^{-\lambda L}$ 
as the number of future predictions $L$, alike the fidelity. Therefore, we plot the rate $\lambda$, that can be 
estimated from the largest eigenvalues of the maps $\mathcal E_{T}$ introduced in Sec.~\ref{sec:compress cl},
which is independent on the memory state.
We call discrepancy the quantity $-\log B$, which is different from zero when $B\neq 1$. 
As shown in Figs.~\ref{fig:qvsc32}(a,b), the compressed quantum simulator has an almost zero discrepancy 
for large $M$, and a negligible decay rate for $M\gtrsim 12$, 
while the classical simulator always displays a larger discrepancy and decay rate, which both start to deviate from zero already at 
$M=31$. In Fig.~\ref{fig:qvsc32}(c) we compare $\lambda$ with the entropy of the steady state, showing in both
cases an increase of the decay rate for reduced entropy. For quantum simulators, there is a significant 
increase when the memory has approximately 1 bit of information, while for classical simulators, that 
normally display a higher entropy, there is a plateau followed by a decrease when the memory has more than 
4 bits. 
From the above analysis we can conclude that, at least for the considered 
compression techniques, the quantum advantage in simulating the discrete renewal process 
persists even when the memory dimension is not enough for exact simulation.

\subsection{Learning data sequences} 
In the previous section we have shown that quantum simulators have the potential to model 
stochastic processes with higher accuracy for a given memory or, in other terms, reach the same 
accuracy with less memory. This outcome was obtained by employing tensor network 
inspired compression methods applied to exact quantum and classical simulators, and is therefore 
algorithmic-dependent. 
%In other terms, we cannot rule out the possibility that the best, yet unknown, classical compression algorithm might outperform the best quantum compression algorithm. 

In this section we test the performance of quantum and classical algorithms on real datasets, namely 
starting from a training data sequence, and learning the optimal classical and quantum simulators. 
In the classical case, we employ the standard expectation-maximization algorithm (Baum-Welch) 
\cite{murphy2012machine}, while in the quantum case we adapt gradient based techniques
\cite{adhikary2019learning,banchi2020convex} to work with exponentially small probabilities. 
We focus on the log-likelihood cost function \cite{glasser2019expressive,yang2023provably}
\begin{equation}
	\mathcal L = -\log P_K(x_1,\dots,x_L), 
	\label{eq:ll}
\end{equation}
where $P_K$ was defined in Eq.~\eqref{eq:prob K} and $x_{1:L}$ defines the training set of $L$ past observations. 
For simplicity,
we set the initial memory state as $\ket\phi=\ket 0$. When $L$ is large, the probabilities $P_K$ 
can be extremely small and cause numerical instabilities and overflows. To fix
this, we note that the direction $G^x$ of 
the steepest descent can be expressed as 
\begin{align}
	G^x &= \sum_{ij} \ket{i}\!\bra j \frac{\partial \mathcal L}{\partial K^{x*}_{ij}} = 
	\sum_{ij} \ket{i}\!\bra j \frac{\partial \mathcal L}{\partial (K^{x\dagger})_{ji}} = 
		\\&= -\sum_{ij} \ket{i}\bra{j} \sum_{t=1}^L \delta_{x,x_t} \times \\&\times\nonumber \frac{
			\bra\phi K^{x_1\dagger}\cdots K^{x_{t-1}\dagger}\ket{j}\!\bra{i} K^{x_{t+1}\dagger} K^{x_T\dagger}
			K^{x_T}\cdots K^{x_1}\ket\phi}{\bra\phi K^{x_1\dagger}\cdots K^{x_T\dagger}K^{x_T\dagger}\cdots K^{x_1}\ket\phi}
			\\& = -
			\sum_{t=1}^L \delta_{x,x_t} \frac{\ket{B_t}\bra{F_t}}{\bra{F_t}K^{x_t\dagger}\ket{B_t}},
			\label{eq:grad}
\end{align}
where we used that 
$\mathcal L$ is a real function of complex matrices $K^{x}$, so the direction of the 
steepest descent corresponds the gradient of the complex conjugate \cite{hjorungnes2007complex,adhikary2019learning},
and, inspired by the classical Baum-Welch algorithm, we have defined the ``forward'' and ``backward'' states 
\begin{align}
	\ket{F_{t+1}} &= \frac{K^{x_t}\ket{F_t}}{\|K^{x_t}\ket{F_t}\|}, & \ket{F_1}&=\ket\phi, \\
	\ket{B_{t-1}} &= \frac{K^{x_t\dagger}\ket{B_t}}{\|K^{x_t\dagger}\ket{B_t}\|}, & \ket{B_1} &= \ket{F_{L+1}}.
\end{align}
Since the above states are iteratively normalized, the numerical instabilities due to the small probabilities are 
removed, and all terms in \eqref{eq:grad} are numerically well-defined. 
From the normalization coefficients we can also write the log-likelihood as 
\begin{equation}
	\mathcal L = -2\sum_{t=1}^{L}\log\|K^{x_t}\ket{F_t}\|,
\end{equation}
to check for convergence, without having to deal with exponentially small probabilities as in Eq.~\eqref{eq:ll}. 

By naively following the steepest descent direction, the updated Kraus operators do not 
define a completely positive map. There are two possible solutions to this problem. The simplest one 
involves the conjugate gradient method \cite{banchi2020convex}, where after updating the parameters as 
$K^x\to K^x+\eta G^x$ with a suitably small learning rate $\eta$, the new states are 
normalized to satisfy $\sum_x K^{x\dagger} K^x=\openone$. In practice, here we use the {\sc mpsnormalize} 
function from Algorithm~\ref{alg:normalize}. For the second approach, we note 
that, by concatenating the Kraus  operators $\mathbb K=[K^1 \dots K^d]$ into a $dD\times D$ matrix
$\mathbb K$, the latter satisfies $\mathbb K^\dagger \mathbb K=\openone$. 
A single $\mathbb K$ is therefore an isometry, and 
the set of isometries define the Stiefel manifold \cite{wen2013feasible}. Using tools from Riemannian geometry, 
the update rule can be defined 
as a curve $\mathbb K(\eta)$ on the Stiefel manifold, which starts from $\mathbb K$ and follows the 
direction defined by the projection of $G^x$ onto the tangent space around $\mathbb K$. 
Although the are several choices  for
the above curve, here we follow the recipe by Wen and Yin \cite{wen2013feasible} and define 
\begin{equation}
	\mathbb K(\eta) = (\openone+\eta \mathbb A)^{-1}(\openone-\eta\mathbb A)\mathbb K,
\end{equation}
where 
\begin{equation}
	\mathbb A = \mathbb G\mathbb K^\dagger - \mathbb K \mathbb G^\dagger,
\end{equation}
and $\mathbb G=[G^1 \dots G^d]$. Note that, when $d$ is large, it is convenient
to manipulate the above formulae to make them depend on smaller matrices
\cite{wen2013feasible}.

\begin{figure}[t]
	\centering
	\includegraphics[width=0.4\textwidth]{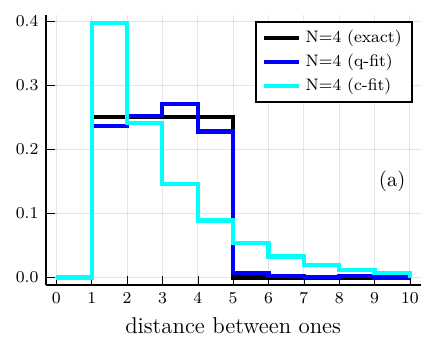}
	\includegraphics[width=0.4\textwidth]{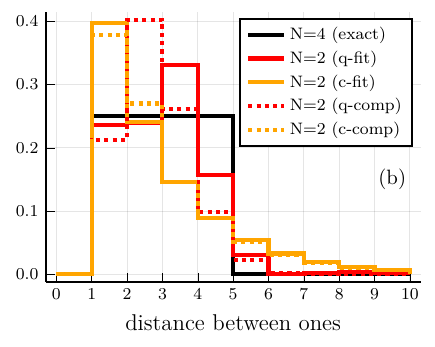}
	\begin{tabular}{|c|cc|} 
		\hline
	&	~~quantum~~ & ~~classical~~ \\
		\hline
		~~$N=4$ (fit)~~ & 0.969  & 0.630 \\ 
		~~$N=2$ (fit)~~ &0.898 & 0.630	\\ 
		~~$N=2$ (comp)~~ & 0.842 & 0.652 \\
		\hline
	\end{tabular}~~~(c)
	\caption{Distribution of the distance between consecutive ones in the exact and 
		fitted classical (c-fit) and quantum (q-fit) simulators, and in 
		the compressed classical (c-comp) and quantum (q-comp) simulators. The exact simulation uses 
		$N=4$, while the fitted ones either $N=4$ (a) or $N=2$ (b). Fitting is done 
		using a training sequence of $L=10^4$ observations, while the histogram is generated using 
		100 million samples. Table (c) shows the Bhattacharyya coefficients of the resulting 
		quantum and classical simulators, either fitted from data or compressed from the exact ones, 
		in predicting the next 10 future observations. 
	}
	\label{fig:qcsamples}
\end{figure}

Numerical results are displayed in Fig.~\ref{fig:qcsamples} and show the superior performance of 
quantum simulators, confirming similar results observed  for 
other stochastic processes \cite{glasser2019expressive,yang2023provably}. 
In particular, in Fig.~\ref{fig:qcsamples}(a), where the fitted classical and quantum models 
have the same memory dimension of the exact simulator, we see that the classical fit 
has long tails, while the quantum one reproduces the expected behaviour, though with 
some fluctuations. In Fig.~\ref{fig:qcsamples}(b), where the memory dimensions of the learnt 
classical and quantum models are constrained to be half that of the exact simulator, we see 
that the differences are less dramatic, though both classical simulators, e.g.~the fitted one and 
the compressed one, keep having longer tails and overall display a larger deviation from 
the true distribution. These larger deviations are summarized in Fig.~\ref{fig:qcsamples}(c),
where we observe that the Bhattacharyya coefficients of the quantum simulators are always 
significantly higher. 
%However, when trying to predict future observations, the learnt classical and quantum models both perform significantly worse than the exact and compressed simulators, namely those obtained from the techniques of Sec.~\ref{sec:compress}.  In other terms, training is likely over-fitting the data, without learning the relevant features of the distribution to predict future outcomes. 
We expect that better generalization capabilities, namely better abilities to predict the future 
given the past, can be
obtained by introducing regularization terms in the cost 
function, e.g.~based on the information bottleneck \cite{banchi2021generalization,banchi2023statistical}. 
These regularization terms should enforce a better exploitation of the memory, with the ultimate aim of letting 
the learnt model save the relevant features from the history of past observations to predict future outcomes.

\begin{table}[t]
	\centering
	\begin{tabular}{|c|cc|}
		\hline
		~~~$D$~~~	& ~~~Quantum~~~ & ~~~Classical~~~ \\ 
		\hline
		16 & 0.037 (0.025) & 0.023 (0.0039) \\
		32 & 0.134 (0.098) & 0.051 (0.0002) \\
		64 & 0.330 (0.261) & 0.002 (0.0002) \\
		%64 & 
		\hline
	\end{tabular}
	\caption{Probability of observing the last 10 outcomes in the Lymphography dataset \cite{misc_lymphography_63},
		given the previous observations, for fitted quantum and classical models. Each table entry shows the 
		best value over 100 repetitions of the fitting procedure for random initial configurations, and 
		the mendian in paranthesis. 
	}
	\label{tab:UCI}
\end{table}

Finally, we apply the fitting technique to real-world data, using the
Lymphography dataset \cite{misc_lymphography_63}.
We fit the lymphatics feature using either classical or quantum models with different dimensions 
and use the model to predict the last 10 values given the history of past observations. 
Since gradient descent converges to a local optimum, we repeat each numerical experiment 100 times with 
different random configurations. The results are shown in Table~\ref{tab:UCI} where 
the quantum advantage is clear. In particular, the quantum models display consistent results 
while the classical ones often converged to poor optima, as shown by the low value of the median. 
For $D=64$, the classical models always converged to poor optima, while for quantum models increasing 
$D$ always increases both the best and median values. 

Code to reproduce the results is provided in \cite{Banchi_Quantum_Tensor_network}.

\section{Conclusions}\label{sec:conclusions}

We have studied whether the quantum advantage in the simulation of classical
stochastic processes persists even when we relax the assumption of exact
simulation, which is unlikely to be reachable with real-world datasets. 

We focused on the trade-off between accuracy and asymptotic memory storage.
We introduced different compression algorithms for classical and quantum
simulators of one-dimensional (e.g.~temporal) stochastic processes, inspired by
tensor network techniques, and analyzed  different figures of merit to assess
how the compressed model deviates from the ideal one when trying to make
predictions. We found that quantum simulators can display a higher accuracy, in
the prediction, for a given memory, or that, alternatively, can achieve the same
accuracy by retaining less information from the training sequence. 

As for future prospects, given that
the information available in the training data bounds the generalization error
\cite{banchi2021generalization}, it is tempting to expect that quantum
simulators may be able to learn a model with less data. A detailed study about
this possibility is left to future investigations. 
%However, we found that simple learning algorithms (e.g.~based on maximum-likelihood), while confirming the predicted advantage of quantum models, fail to reach the remarkable accuracy of the models obtained by compressing a known classical or quantum simulator. Therefore, more powerful learning algorithms need to be developed in order to fully exploit the advantage of quantum models for simulating classical stochastic processes. 
Another future prospect concerns the use of quantum hardware capable of 
implementing mid-circuit measurements \cite{deist2022mid}. Classical tensor 
network simulations are limited to matrices with a small memory dimension, while 
quantum computers are capable of simulating circuits like the one in Fig.~\ref{fig:cartoon} 
with a memory dimension that increases exponentially with the number of memory qubits. 
However, unlike tensor network simulations, the unitary operations are less general 
and constrained, e.g., to shallow quantum circuits. Furthermore, the overall depth 
of the circuit should be small, though longer depths can be obtained with circuit-cutting 
techniques \cite{lowe2023fast}. It is therefore interesting to 
explore the possibility of using a quantum hardware for two reasons: 
i) to see whether the quantum advantage in memory use persists even with the above constraints, and 
ii) to explore the possibility of the computational advantage enabled by quantum algorithms. 

\begin{acknowledgements}
This work was supported by the European Union
under the Italian National Recovery and Resilience Plan
(NRRP) of NextGenerationEU, partnership on
``Telecommunications of the Future'' (PE00000001 - program
``RESTART''). 
\end{acknowledgements}

\appendix

\section{Non-normalized quantum simulators}\label{app:nonorm}

Here we make some remarks on non-normalized quantum simulators 
with tensor $A$. As shown in Algorithm~\ref{alg:normalize}, from $A$ 
we can define a new tensor $B$ with elements $B^x= WA^xW^{-1} /\sqrt\mu$, 
where $W=\sqrt s U^\dagger$. Note that the dominant eigenvalue 
$\mu$ is real and positive \cite{evans1977spectral}.
When $A$ is not-normalized, Eq.\eqref{eq:P unifilar} does not 
define a normalized probability probability distribution. In order 
to fix this issue, for non-normalized simulators we multiply 
Eq.~\eqref{eq:quantum from A} with a left operator $O$ and write 
\begin{align}
	\ket{\psi_S(x_1,\dots,x_L)} &= \frac{OA^{x_L}\cdots A^{x_1}\ket{\sigma_A}}{P_A(x_1,\dots,x_L)}
	\label{eq:quantum from A nn}
	\\
%	A^x\ket{\sigma_\beta} & = \sum_\alpha \sqrt{T^x_{\alpha\beta}} \ket{\sigma_\alpha}, \\ 
	P_A(x_1,\dots,x_L) &= \|OA^{x_L}\cdots A^{x_1}\ket{\sigma_A}\|^2_2.
	\label{eq:prob nn}
\end{align}
From the tensor $B$ we may define
$	P_B(x_1,\dots,x_L) = \|B^{x_L}\cdots B^{x_1}\ket{\sigma_B}\|^2_2$,
where the left operator is not required, since $B$ defines  normalized 
Kraus operators. Moreover, since $W$ and $W^{-1}$ cancel after each 
iteration 
\begin{equation}
P_A(x_1,\dots,x_L) = \mu^L\|OW^{-1}B^{x_L}\cdots B^{x_1}W\ket{\sigma_A}\|_2^2,
\nonumber
\end{equation}
so $\sum_{\{x\}} P_A(x_1,\dots,x_L)=1$ if 
\begin{equation}
	O = \mu^{-L}\frac{W}{\sqrt{\bra{\sigma_A}W^\dagger W\ket{\sigma_A}}}.
\end{equation}
With the above choice, and assuming that $\ket{\sigma_B}$ is normalized, 
the two probability distributions become identical 
$P_A(x_1,\dots,x_L)=P_B(x_1,\dots,x_L)$ as long as 
\begin{equation}
\ket{\sigma_A}= \frac{W^{-1}\ket{\sigma_B}}{\sqrt{\bra{\sigma_B}(W^{-1})^{\dagger} W^{-1}\ket{\sigma_B}}}.	
\end{equation}

\section{Quantum Simulation of Non-Unifilar HMMs}\label{a:not unifilar}
Beyond the unifilar case, unitary simulation of an HMM with transition tensor $T$ is 
possible by using two ancillary states \cite{elliott2021memory} 
\begin{equation}
U\ket{0,\sigma_\beta,0,0} = \sum_{x\alpha} \sqrt{T^x_{\alpha\beta}} \ket{x,\sigma_\alpha,\alpha,x}.
\end{equation}
Tracing out the extra two degrees of freedom we note that this unitary simulator, in 
place of Eq.~\eqref{eq:quantum from A} yields 
a matrix product density operator
\begin{equation}
	\rho = \sum_{\{x_t\}} \rho_S(x_1,\dots,x_L) \ket{x_1,\dots,x_L}\bra{x_1,\dots,x_L},
\end{equation}
with 
\begin{align}
	\rho_S(x_1,\dots,x_L) &= \frac{\mathcal E_T^{x_L}\cdots \mathcal E_T^{x_1}(\rho_0)}{
%	P_T(x_1,\dots,x_L)
%	\\
%	P_T(x_1,\dots,x_L) &= 
		\Tr[\mathcal E_T^{x_L}\cdots \mathcal E_T^{x_1}(\rho_0)]
} ,
\end{align}
where $\mathcal E_T^x[\ket{\sigma_{\beta}}\bra{\sigma_\beta}] = \sum_{x,\alpha} T^x_{\alpha\beta} 
\ket{\sigma_{\alpha}}\bra{\sigma_\alpha}$ is a classical channel.

\section{Spectral Compression of Hidden Markov Models}\label{a:cl compress}

\begin{figure}[t]
\begin{algorithm}[H]
	\caption{Compress a Hidden Markov Model} 
  \label{alg:classical compress a}
   \begin{algorithmic}[1]
		 \Require{ A normalized $d\times D\times D$ transition tensor  $T$ and the
		 dimension $D'<D$ of the compressed memory}
		 \Function{hmmcompress}{$T,D'$}
		 \State Compute the transition matrix $J_{\alpha\beta}= \sum_x T^{x}_{\alpha\beta}$, 
		 \StatexIndent[1.5] 
		 the emission matrix $E^x_\beta = \sum_\alpha T^{x}_{\alpha\beta}$, the steady state 
		 \StatexIndent[1.5] 
		 $J\pi=\pi$ and the ``frequency'' matrix $F=J\rm{diag}(\pi)$. 
		 \State Compute the singular value decomposition 
		 \StatexIndent[1.5] 
		 $F=UsV^T$, with singular values 
		 in decreasing order. 
		 \State Define $B^x_{\bar\alpha\bar\beta} = \sum_{\gamma=1}^D E^x_\gamma Q_{\bar\alpha\gamma}
		 U_{\gamma\bar\beta}$, where 
		 \StatexIndent[1.5] $Q={\rm diag}(s)V^T$ and
		 $\bar\alpha,\bar\beta=1,\dots,D'$.
		 \ForAll {$\bar\alpha,\bar\beta,x$ such that  $B^x_{\bar\alpha\bar\beta}\leq 0$}
		 \State set $B^x_{\bar\alpha\bar\beta} =0$
		 \EndFor
		 \State Compute the normalization factors $N_{\bar\beta} = \sum_{x\bar\alpha} B^x_{\bar\alpha\bar\beta}$
		 \State Set $\bar T^x_{\bar\alpha\bar\beta}=B^x_{\bar\alpha\bar\beta}/N_{\bar\beta}$ if 
		 $N_{\bar\beta}\neq0$ or $\bar T^x_{\bar\alpha\bar\beta}=1/(dD')$ 
		 \StatexIndent[1.5] 
		 otherwise. 
		 \State \Return $\bar T^x_{\bar\alpha\bar\beta}$
		 \EndFunction
   \end{algorithmic}
\end{algorithm}
\end{figure}
We focus on 
adapting spectral compression methods \cite{zhang2019spectral,wu2010probability}, which
share similar ideas with Algorithm~\ref{alg:quantum compress}, though they 
were developed for reducing the rank of the transition matrix, rather than 
the memory dimension. However, these two tasks are related.  
Consider a Markov transition matrix $T_{\alpha,\beta}=P(s_t{=}\alpha|s_{t-1}{=}\beta)$, 
then a low-rank approximation is given by 
\begin{equation}
T_{\alpha,\beta} \approx \sum_{\zeta=1}^r P(s_{t}{=}\alpha|z{=}\zeta)P(z{=}\zeta|s_{t-1})
\end{equation}
where $z$ is a new stochastic variable taking values $1,\dots,r$, where $r$ is the desired 
compressed rank. However, it turns out that finding such decomposition is rather 
complicated, so Ref.~\cite{zhang2019spectral} proposes to decompose $T$ via 
the singular value decomposition, and then project the resulting matrix to the space of 
properly normalized transition matrices. Inspired by such technique, we 
introduce Algorithm~\ref{alg:classical compress a}. The main difference between
Algorithm~\ref{alg:classical compress a} and that of Ref.~\cite{zhang2019spectral} is that 
we get a HMM acting on a reduced memory, rather than a HMM acting on the same memory, but with 
lower rank. In order to do so, we use the matrix product form \eqref{eq:P mps} to 
shift the $U$ matrix from the singular value decomposition to the right and only then perform a 
projection to get a proper transition tensor.

%Applying such decomposition to the (rectangular) matrix \eqref{eq:P def} we get  $T^x_{\alpha\beta} \approx \sum_{\zeta=1}^r M^x_{\alpha\zeta} Q_{\zeta\beta}$ where both $M$ and $Q$ define conditional probabilities, so the positivity and normalization of $T^x_{\alpha\beta}$ is preserved. In order to defined a compressed memory, we notice that in the matrix product expression \eqref{eq:P mps} we can swap the orders of $M$ and $Q$ in order to get an identical expression, but acting on a reduced memory 
%\begin{equation}
%	T^{x}_{\alpha\beta} 
%	~~~~~\longrightarrow~~~~~
%	\bar{T}^x_{\bar\alpha\bar\beta} = \sum_\alpha Q_{\bar\alpha\alpha}M^x_{\alpha\bar\beta}
%\end{equation}
%where $\bar\alpha,\bar\beta=1,\dots,r$, $\bar{T}^x_{\bar\alpha\bar\beta}\geq 0$ and $\sum_{x,\bar\alpha}\bar{T}^x_{\bar\alpha\bar\beta}=1$. 


\end{document}